\documentclass[11pt,fleqn,twoside]{article}
\usepackage{amsfonts,latexsym}
\makeatletter
\newcommand{\prava}{\footnotesize\it
\begin{flushright}
\begin{minipage}{18cm}
Copyright \copyright 1998 by K. Porsezian
\end{minipage}
\end{flushright}}

\newcommand{\name}[1]{\begin{flushleft}
                       \LARGE \bf #1
                       \end{flushleft}\vspace{-3mm}}

\newcommand{\Author}[1]{\begin{flushleft}
                       \it #1 \end{flushleft}}

\newcommand{\Adress}[1]{\begin{flushleft}
                       \it #1 \end{flushleft}}

\newcommand{\Date}[1]{\begin{flushleft}
                      \small  \it #1 \end{flushleft}}

\newcommand{\ehkol}{Author \ name}
\newcommand{\ohkol}{Article \ name}
\renewcommand{\@evenhead}{
\hspace*{-3pt}\raisebox{-15pt}[\headheight][0pt]{\vbox{\hbox to \textwidth
{\thepage \hfil \ehkol}\vskip4pt \hrule}}}
\renewcommand{\@oddhead}{
\hspace*{-3pt}\raisebox{-15pt}[\headheight][0pt]{\vbox{\hbox to \textwidth
{\ohkol \hfil \thepage}\vskip4pt\hrule}}}
\renewcommand{\@evenfoot}{}
\renewcommand{\@oddfoot}{}

\newcommand{\be}{\begin{equation}}
\newcommand{\ee}{\end{equation}}
\newcommand{\ba}{\hspace*{-5pt}\begin{array}}
\newcommand{\ea}{\end{array}}

\newcommand{\ds}{\displaystyle}

\makeatother

\begin{document}
\setcounter{page}{126}

\thispagestyle{empty}

\renewcommand{\ehkol}{K. Porsezian}
\renewcommand{\ohkol}{Bilinearization of Coupled Nonlinear
Schr\"odinger Type Equations}

\begin{flushleft}
\footnotesize \sf
Journal of Nonlinear Mathematical Physics \qquad 1998, V.5, N~2,
\pageref{porsezian-fp}--\pageref{porsezian-lp}. \hfill {\sc Letter}
\end{flushleft}

\vspace{-5mm}

{\renewcommand{\footnoterule}{}
{\renewcommand{\thefootnote}{}  \footnote{\prava}}

\name{Bilinearization of Coupled Nonlinear Schr\"odinger Type
Equations: Integrabilty and Solitons}\label{porsezian-fp}

\Author{K. PORSEZIAN}

\Adress{Institut f\"ur Theoretische Physik, Universit\"at Hannover,\\
30167 Hannover, Germany\\[1mm]
 and \\[1mm]
Centre for Laser Technology, Department of Physics,
Anna University, \\
Madras 600 025, India}

\Date{Received November 19, 1997; Accepted February 23, 1998}

\begin{abstract}
\noindent
Considering the coupled envelope equations in nonlinear couplers, the
question of integrability is attempted. It is explicitly shown that
Hirota's bilinear method is one of the simple and alternative
techniques to Painlev\'e
analysis to obtain the  integrability conditions of the
coupled nonlinear Schr\"odinger (CNLS) type  equations. We also show
that the coupled Hirota equation introduced by Tasgal
and Potasek is the next hierarchy of the inverse scattering solvable CNLS
equation.  The results are in agreement with the known results.
\end{abstract}


\section{Introduction}
The propagation of temporal soliton envelopes in nonlinear optical media has
been predicted and demonstrated experimentally
\cite{porsezian:AB1}--\cite{porsezian:AB5}.
This
prediction arises from the opportunity of reducing the Maxwell equations
which govern the propagation to a single, completely integrable, partial
dif\/ferential equation(PDE) where one is left with the time and the
propagation
distance as independent variables~\cite{porsezian:AB1}. The best and
well studied
example is  the nonlinear Schr\"odinger (NLS) equation
\begin{equation}
iq_z +q_{tt} + 2\mid{q}\mid^2q =0
\end{equation}
which describes the wave propagation of picosecond-pulse envelopes $q(z,t)$ in
a lossless single-mode f\/ibre. Eq.(1) is one of the completely integrable
$N$-soliton possessing
systems. Eventhough eq.(1) may adequately describe the propagation in
a single-mode waveguides, routing-, switching operations, and other optical
ef\/fects which involve soliton pulses require the interaction between two or
more modes
\cite{porsezian:AB4}--\cite{porsezian:AB8}.
In general, the coupled mode approach~\cite{porsezian:AB7} still
permits description of the pulse
propagation in a multi-mode waveguide by means of vector versions of eq.(1).
Although these systems of equations are no longer integrable, one may obtain
quantitative information about the pulse propagation  by resorting to
numerical or perturbative methods. Situations of physical interest that can
be described by CNLS equations include two parallel waveguides
coupled through evansecnt f\/ield overlap, the coupling of two polarizations
modes in uniform guides, and so on. Note that the study of the propagation
of optical solitons in multi-mode nonlinear couplers, besides being important
from the theoretical point of view, is also  important in view of their
possible
applications. Recent advances have permitted proof that solitons
are ideal candidates for performing all-optical switching operations in
nonlinear couplers. In fact, their stability leads to the possibility of
controlling the coupling of the whole pulse, by means of changing the
input power of a single pulse, or the phase of the superimposed pulse.

For the past few years, the wave propagation in nonlinear couplers have been
investigated from theoretical, numerical and experimental point of view
\cite{porsezian:AB4}--\cite{porsezian:AB19}.
From
the detailed theoretical investigations, several integrable soliton
possessing  coupled NLS equations have been
introduced
\cite{porsezian:AB5,porsezian:AB9,porsezian:AB10,porsezian:AB12,
porsezian:AB16,porsezian:AB19}. The experimental results
reveals that these integrable soliton
models are not  accurate enough to explain the wave propagation in nonlinear
couplers. On the otherhand, theoretically several new results and concepts
have been introduced. Thus it is clear that to propagate exact soliton through
nonlinear couplers, one has to look for dif\/ferent conditions among the
 parameters involved in the system. In this letter, we present new type of
bilinear conditions, which may be very useful for the experimental
realization of solitons in nonlinear couplers.
}

It is well known that Painlev\'e singularity structure analysis
is one of the systematic and powerful method in nonlinear science to identify
the integrability conditions of nonlinear partial dif\/ferential  equations
(NPDEs) \cite{porsezian:AB18}. In recent years, this method has been
applied to a very
large number of NPDEs and also systematically estabilished the complete
integrability properties like Lax pair, B\"acklund transformation,
bilinear transformation, soliton solutions, and so
on
\cite{porsezian:AB13,porsezian:AB20,porsezian:AB21,porsezian:AB22}.
For CNLS type
equations, this method is found to be cumbersome and the construction of the
Lax pair from this analysis is still an open question. Also, after getting the
 necessary condition(s) for integrability, one has to apply bilinear or other
methods to prove the suf\/f\/icient condition for the complete
integrability. The
main aim of this letter is to show explicitly  that  Hirota's bilinear
approach \cite{porsezian:AB23} can be used as an alternative method to
Painlev\'e analysis to identify  the complete and partial
integrability conditions of CNLS equations with many parameters.  Using
this method, we show that  one can obtain the more general bilinear
(integrability) conditions of coupled nonlinear equations in a very
simple way.

\section{Two Coupled Nonlinear Schr\"odinger Equation}
There are several methods to derive  a set of CNLS equations, depending on
the physical situation that is being modeled. A fairly general and
frequently studied CNLS equation is of the
form \cite{porsezian:AB4,porsezian:AB5,porsezian:AB6,porsezian:AB8,porsezian:AB10,porsezian:AB12,porsezian:AB13}
\begin{equation}
iu_z = c_1u_{tt}+2\left(\alpha \mid u \mid^2 +\beta \mid v \mid^2\right)u,
\end{equation}
\begin{equation}
iv_z = c_2v_{tt}+2\left(\beta \mid u \mid^2 +\gamma \mid v\mid^2\right)v.
\end{equation}
Eqs.(2-3) are shown to be completely  integarble for the following two cases:
\begin{equation}
(i): \ c_1 = c_2  , \quad \alpha = \beta = \gamma;\qquad (ii): \
c_1 = -c_2, \quad \alpha = -\beta = \gamma.
\end{equation}
To obtain the (bilinear)integrability conditions and soliton solutions,
we introduce the following transformation \cite{porsezian:AB23}
\begin{equation}
u = \frac{G}{F},\qquad v = \frac{H}{F},
\end{equation}
where $G$ and $H$ are complex functions and $F$ is a real function
and the bilinear operator is def\/ined by
\begin{equation}
D_t^m D_z^n = \left(\frac{\partial}{\partial{t}}-
\frac{\partial}{\partial{t'}}\right)^m
\left(\frac{\partial}{\partial{z}}- \frac{\partial}{\partial{z'}}\right)^n
G(z,t)F(z',t')
\mid z'=z, \ t'=t.
\end{equation}
Substituting eq.(5) in eqs(2-3), we obtain
\begin{equation}
\left(iD_z-c_1D_t^2 \right)G\cdot F=0,\qquad
\left(iD_z-c_2D_t^2 \right)H\cdot F=0,
\end{equation}
\begin{equation}
D_t^2 F\cdot F = \frac{2}{c_1}\left(\alpha GG^*+\beta HH^*\right),
\qquad D_t^2 F\cdot F = \frac{2}{c_2}\left(\gamma HH^*+\beta GG^*\right).
\end{equation}
So the left hand sides of eq.(8) become equal. Hence the right hand sides of
eq.(8) should also be equal which is true only when
\begin{equation}
c_2\alpha = c_1\beta , \qquad c_2\beta = c_1\gamma.
\end{equation}
The above conditions can be obtained by equating the coef\/f\/icients
of $GG^*$
and $HH^*$ respectively in eq.(8). One can easily check that  eq.(9)
admits the already known integrability conditions (4). Solving eq.(9),
we obtain
\begin{equation}
c_2 = \pm \sqrt{\frac{\gamma}{\alpha}} c_1 , \qquad \alpha\gamma  = \beta^2.
\end{equation}
It should be noted that conditions (10) are more general than (4). As the
bilinear conditions (9) or (10) are in terms of all parameters, we
believe that these conditions may be very useful for the experimental
generation of solitons  in nonlinear couplers.

Having obtained the bilinear conditions and bilinear forms, our next
aim is to obtain the soliton solutions. Now to generate the soliton solutions,
we expand the dependent variables in terms of $\varepsilon$ and the solutions
can be constructed as usual. For  instance, if $G = \varepsilon G_1 ,H =
\varepsilon H_1$ and $F = 1+\varepsilon^2F_2$ , the solutions are found to be
\begin{equation}
G_1 = e^{\eta_1}, \qquad H = e^{\eta_2 +\xi} , \qquad F_2 =
\frac{\beta e^{\eta_2+\eta_2^*
+\xi+\xi^*} +\alpha e^{(\eta_1 +\eta_1^*)}}{c_1(k_1 +k_1^*)^2},
\end{equation}
where $\eta_1 = k_1(t-ic_1k_1z) + \eta_1^0$, $\eta_2 = k_1(t-ic_2k_1z)
+\eta_1^0$ and $\xi$ is a complex constant. From the detailed investigations,
 we f\/ind that eqs.(2-3) admit the higher order soliton solutions for
the conditions given in eq.(4) and the resultant solutions are  in agreement
with the resluts reported earlier
\cite{porsezian:AB9,porsezian:AB14,porsezian:AB15}. So like f\/ixing
the parameters in the arbitrary analysis of Painlev\'e method, using
bilinear method one can also get the complete integrability conditions from
the generation of higher order soliton solutions.

\section{Three Coupled NLS Equations}

Considering the systems of three coupled NLS equations of the
form~\cite{porsezian:AB14}
\begin{equation}
iu_z =c_1u_{tt}+2\left(\alpha \mid u \mid^2 +\beta \mid v \mid^2 +
\delta\mid w \mid ^2\right)u,
\end{equation}
\begin{equation}
iv_z = c_2v_{tt}+2\left(\beta \mid u \mid^2 +\gamma \mid v\mid^2+
\Gamma\mid w \mid ^2\right)v,
\end{equation}
\begin{equation}
iw_z = c_3w_{tt}+2\left(\delta\mid u \mid^2 +\Gamma \mid v\mid^2+\Delta\mid w
\mid ^2\right)w.
\end{equation}
Using the similar procedure the bilinear conditions are found to be of
the form:
\begin{equation}
  c_2c_3\alpha = c_1c_3\beta = c_1c_2\delta,
\quad c_2c_3\beta = c_1c_3\gamma= c_1c_2\Gamma,
\quad c_2c_3\delta = c_1c_3\gamma = c_1c_2\Delta.
\end{equation}

Here also we f\/ind that eqs.(15) satisfy the integrability
conditions reported through Painlev\'e analysis~\cite{porsezian:AB14} and
the soliton solutions can also be constructed.

\section{Coupled Hirota equation}

In this section, we will discuss the integrability of the coupled Hirota
equation~\cite{porsezian:AB16}.
In order to increase the bit rates  it is necessary to decrease the pulse
width of the order of femtosecond. As pulse lengths become comparable to the
wavelength, however, eqs.(2-3) are inadequate, as additional
ef\/fects must now
be considered. By considering the above facts, a generalized coupled Hirota
equation  reads as~\cite{porsezian:AB16}
\begin{equation}
\ba{l}
iu_z +c_1u_{tt}+2\left(\alpha \mid u \mid^2 +\beta \mid v
\mid^2\right)u\\[2mm]
\qquad -i\epsilon \left[u_{ttt} +\left(2\mu_1\mid{u}\mid^2
+\nu_1\mid{v}\mid^2\right)u_t +\nu_1 uv^*v_t\right] = 0,
\ea
\end{equation}
\begin{equation}
\ba{l}
iv_z + c_2v_{tt}+2\left(\beta \mid u \mid^2 +\gamma \mid
v\mid^2\right)v\\[2mm]
\qquad -i\epsilon \left[v_{ttt} +\left(\nu_2\mid{u}\mid^2
+2\mu_2\mid{v}\mid^2\right)v_t +\nu_2 u^*vu_t\right] = 0.
\ea
\end{equation}
When $\ds c_1 = c_2 = \frac{\lambda}{2}$, $\alpha =\beta =\gamma = \lambda$,
$\mu_1 =\nu_1 = \mu_2 =\nu_2 = 3$,
eqs.(16-17) take the form of the equations in \cite{porsezian:AB16}.
In eqs.(16-17), we have not included the linear ef\/fects, which can
always be removed through suitable transformation in $u$ and $v$
\cite{porsezian:AB15} and,
for our convenience, we introduce the new parameters $\nu_2$ and $\mu_2$. The
bilinear forms of (16-17) are in the form
\begin{equation}
\left(iD_z+c_1D_t^2 -i\epsilon D_t^3\right)G\cdot F=0,
\qquad \left(iD_z+c_2D_t^2
-i\epsilon D_t^3 \right)H\cdot F=0,
\end{equation}
\begin{equation}
D_t^2 F\cdot F = \frac{2}{c_1}\left(\alpha GG^*+\beta HH^*\right),
\qquad
D_t^2 F\cdot F = \frac{2}{3}\left(\mu_1 GG^*+\nu_1 HH^*\right),
\end{equation}
\begin{equation}
D_t^2 F\cdot F = \frac{2}{c_2}\left(\gamma HH^*+\beta GG^*\right),
\qquad
D_t^2 F\cdot F = \frac{2}{3}\left(\nu_2 GG^*+\mu_2 HH^*\right),
\end{equation}
The more general bilinear conditions are:
\begin{equation}
\frac{\alpha}{c_1} = \frac{\mu_1}{3} = \frac{\beta}{c_2} =
\frac{\nu_2}{3},
\qquad \frac{\beta}{c_1} = \frac{\nu_1}{3} = \frac{\gamma}{c_2} = \frac{\mu_2}{3}.
\end{equation}
Here also we f\/ind that  eq.(21)  admits the higher order soliton  solutions
for the conditions $c_1 = c_2$ , $\alpha = \beta = \gamma$, and $\mu_1
=\nu_1 = \mu_2 =\nu_2= 3$ \cite{porsezian:AB16,porsezian:AB17}. Exact
$N$-envelope soliton solutions
of eqs.(16-17) can also be expressed in a similar  way.  It is interesting to
note that, like CNLS, eqs.(16-17) also found to be completely integrable for
another conditions $c_1 = -c_2$ , $\alpha =- \beta = \gamma$, and
$\mu_1 = -\nu_1= -\mu_2 = \nu_2=  3$.
It should be noted that the latter condition is one of the new integrable
system obtained through this method. Recently, using multi-Hamiltonian
formalism, we also verif\/ied the above complete integrability conditions and
found that the above integrable cases are the next hierarchy of the CNLS
equations with conditions (4) \cite{porsezian:AB18}. Hence, the
coupled Hirota equation
introduced by Tasgal and Potasek~\cite{porsezian:AB16} is the next
hierarchy of the IST solvable CNLS equation-I.

\section{Coupled Higher Order NLS Equation}

We consider the coupled higher order NLS equation of the
form~\cite{porsezian:AB19}
\begin{equation}
\ba{l}
iu_z +c_1u_{tt}+2\left(\alpha \mid u \mid^2 +\beta \mid v
\mid^2\right)u\\[2mm]
\qquad -i\varepsilon \left[u_{ttt} +6\left(\mid{u}\mid^2
+\mid{v}\mid^2\right)u_t
+3\left(\mid{u}\mid^2+\mid{v}\mid^2\right)_tu \right] = 0,
\ea
\end{equation}
\begin{equation}
\ba{l}
iv_z + c_2v_{tt}+2\left(\beta \mid u \mid^2 +\gamma \mid
v\mid^2\right)v\\[2mm]
\qquad -i\varepsilon \left[v_{ttt} +6\left(\mid{u}\mid^2
+\mid{v}\mid^2\right)v_t
+3\left(\mid{u}\mid^2+\mid{v}\mid^2\right)_tv\right] = 0.
\ea
\end{equation}
Equations (22-23) can be derived from the higher order NLS
\cite{porsezian:AB3,porsezian:AB5} by
considering the electromagnetic wave $\vec{E}$ as a sum of left and right
polarized waves~\cite{porsezian:AB19}. When $\ds c_1 = c_2 = \frac{1}{2}$
and $\alpha = \beta =\gamma =1 $, the bilinear forms  are  found to be
\begin{equation}
\left(iD_z+\frac{1}{2}D_t^2 -i\varepsilon D_t^3\right)G\cdot F=0,
\quad \left(iD_z+\frac{1}{2}D_t^2
-i\varepsilon D_t^3 \right)H\cdot F=0,
\end{equation}
\begin{equation}
D_t^2 F\cdot F = 4(GG^*+HH^*),
\end{equation}
\begin{equation}
D_tG\cdot G^* = 0,  \quad D_tG\cdot H^* = 0,\quad
D_tG^*\cdot H = 0, \quad D_tG\cdot H = 0, \quad D_tH\cdot H^* = 0 .
\end{equation}
The one soliton solution of eqs.(22-23) is constructed as
\begin{equation}
u = \frac{a_1k_1}{\sqrt{2(\mid{a_1}\mid^2 +\mid{a_2}\mid^2)}}\mbox{sech}
\left(k_1t+\varepsilon
k_1^3z +\eta_1^0\right)\exp\left(\frac{ik_1^2z}{2}\right),
\end{equation}
\begin{equation}
v = \frac{a_2k_1}{\sqrt{2(\mid{a_1}\mid^2 +\mid{a_2}\mid^2)}}\mbox{sech}
\left(k_1t+\varepsilon
k_1^3z +\eta_1^0\right)\exp\left(\frac{ik_1^2z}{2}\right),
\end{equation}
where $a_1$ and  $a_2$ are integration constants.

Thus in this letter, we have shown that Hirota's bilinear method is one of
the alternative formalism to Painlev\'e singularity structure analysis
to identify the integrability of coupled nonlinear Schr\"odinger
equations with many parameters. This analysis is found to be very simple when
compared with the Painlev\'e analysis. Also from the
optical point of view, the general bilinear conditions obtained through
bilinear form may be
useful for the experimental generation of solitons in nonlinear couplers.
Another advantage of this method is that one can, in a simple manner,
construct
 the soliton solutions for all integrable cases. We have also pointed out that
 the coupled Hirota equation studied by Tasgal and Potasek is  the next
hierarchy of the IST solvable CNLS equation. We 
found that the conditions obtained through this method are in agreement with
the results reported earlier. It will be  interesting to
investigate the nature of the solutions (at least two soliton solutions) for
the new integrable cases. Work is in progress in this
direction.

\medskip

\noindent
{\bf Acknowledgments:}
The author wishes to  thank Professor H.J.~Mikeska for his kind hospitality
and constant encouragement. He expresses his thanks to DAAD for
of\/fering the fellowship and DST and CSIR, Govt of India, for the
f\/inanical support through major research grants.

 \label{porsezian-lp}
\end{document}